\documentstyle[12pt]{article}
\newcommand{\be}[1]{\begin{equation}\label{#1}}
\newcommand{\ba}[1]{\begin{eqnarray}\label{#1}}
\newcommand{\ee}{\end{equation}}
\newcommand{\ea}{\end{eqnarray}}
\newcommand{\non}{\nonumber\\\rule{0pt}{30pt}}

\newcommand{\dis}{\displaystyle}
\newcommand{\eq}[1]{(\ref{#1})}
\newcommand{\stint}{\int_{-\infty}^\infty}
\begin{document}
\begin{center}
{\Large\bf A nonlinear indentity for\\
the scattering phase of integrable models}\\
\vskip1cm
{\large N.~A.~Slavnov}\\
\vskip5mm
{\it Steklov Mathematical Institute,
Gubkina 8, Moscow 117966, Russia.}\\
nslavnov@mi.ras.ru
\begin{abstract}
\noindent
A nonlinear identity for the scattering phase of quantum
integrable models is proved.
\end{abstract}
\end{center}
\vskip1cm

In the paper \cite{KS} a new nonlinear identity for the scattering 
phase of quantum integrable models was found. In the present paper we
give a generalization of this identity and show that the last one is 
a corollary of some simple properties of integral transformations of 
rather general type. The identity, which will be discussed, is valid 
for a wide class of integrable models, however in order to specify our 
formul\ae~we consider a well known example of an one-dimensional Bose 
gas with delta-function interactions.

The main equation, describing the thermodynamics of the 
one-di\-men\-si\-o\-nal Bose gas, is the Yang--Yang equation 
for the energy of an one-particle excitation \cite{YY} 
\be{YY} 
\varepsilon(\lambda)=\lambda^2-h-\frac{T}{2\pi}
\stint K(\lambda-\mu)\ln\left(1+e^{-\varepsilon(\mu)/T}
\right)\,d\mu.
\ee
Here $\lambda$ is a spectral parameter, $c>0$ is a coupling constant,
$h>0$ is a chemical potential, and $T\ge 0$ is a temperature. The
kernel $K(\lambda-\mu)$ is equal to
\be{kernel}
K(\lambda-\mu)=\frac{2c}{(\lambda-\mu)^2+c^2}.
\ee
The equation \eq{YY} is uniquely solvable \cite{YY}. Below
we shall use the following properties of the function $\varepsilon(
\lambda)$: this is an even function $\varepsilon(\lambda)=
\varepsilon(-\lambda)$; it has two real roots  $\varepsilon(\pm 
q_T)=0$, where $q_T>0$; $\varepsilon(\lambda)>0$, if $|\lambda|>q_T$ 
and $\varepsilon(\lambda)<0$, if $|\lambda|<q_T$.

The $S$-matrix of the model, defining the scattering of the
particle with the spectral parameter $\lambda$ on the
particle with the spectral parameter $\mu$, is 
\be{Smatr}
S(\lambda,\mu)=\exp\{2\pi iF(\lambda,\mu)\},
\ee
where the scattering phase $F(\lambda,\mu)$ can be found from the 
integral equation \cite{K}
\be{Tphase}
F(\lambda,\mu)-\frac{1}{2\pi}
\stint K(\lambda-\nu)\vartheta(\nu)F(\nu,\mu)\,d\nu
=\frac{1}{2\pi}\Phi(\lambda-\mu).
\ee
Here $\vartheta(\lambda)$ is the Fermi weight
\be{Fermi}
\vartheta(\lambda)=\left(1+e^{\varepsilon(\lambda)/T}\right)^{-1},
\ee
and $\Phi(\lambda)$ is the bare phase
\be{barephase}
\Phi(\lambda)=i\ln\left(\frac{ic+\lambda}{ic-\lambda}\right).
\ee
Observe, that the kernel of the equations \eq{YY} and \eq{Tphase}
is the derivative of the bare phase
\be{der}
K(\lambda-\mu)=\Phi'(\lambda-\mu).
\ee
The equation \eq{Tphase} also is uniquely solvable.

In the limit $T\to 0$ (ground state \cite{L1}, \cite{L2}) the 
equation \eq{Tphase} simplifies. Indeed, in this limit the Fermi 
weight $\vartheta(\lambda)$ goes to the Heaviside step function 
$\vartheta(\lambda)\to\theta(q^2-\lambda^2)$, where
$q=\lim\limits_{T\to 0}q_T$ is the value of the spectral parameter 
on the boundary of the Fermi sphere. The equation \eq{Tphase} turns 
into
\be{0phase}
F_0(\lambda,\mu)-\frac{1}{2\pi}
\int_{-q}^q K(\lambda-\nu)F_0(\nu,\mu)\,d\nu
=\frac{1}{2\pi}\Phi(\lambda-\mu),
\ee
where $F_0(\lambda,\mu)=\lim\limits_{T\to 0}F(\lambda,\mu)$.
The identity, found in \cite{KS}, relates the values of the 
scattering phase on the boundaries of the Fermi sphere
\be{ident0}
\Bigl(1-F_0(q,q)\Bigr)\Bigl(1+F_0(-q,-q)\Bigr)+
F_0(q,-q)F_0(-q,q)=1.
\ee
Such the nonlinear identity for the solution of the linear integral 
equation \eq{0phase} looks rather unsuspected. However, this is on 
particular case of more general identity, which is valid for the 
scattering phase at finite temperature
\be{ident1}
F(\lambda,\mu)-F(-\mu,-\lambda)=
\stint F(\lambda,\nu)F(-\mu,-\nu)\partial_\nu \vartheta(\nu)
\,d\nu.
\ee 
It is easy to see that in the limit $T\to0$ the derivative of the 
Fermi weight  results into the linear combination of delta-functions 
$\partial_\lambda \vartheta(\lambda)=\delta(\lambda+q)
-\delta(\lambda-q)$, hence the identity \eq{ident1} implies 
\be{ident2}
F_0(\lambda,\mu)-F_0(-\mu,-\lambda)=
F_0(\lambda,-q)F_0(-\mu,q)-F_0(\lambda,q)F_0(-\mu,-q).
\ee 
Finally, putting $\lambda=\mu=q$,  we arrive at \eq{ident0}. 

As for the proof of the main identity \eq{ident1}, it is based on rather 
simple observation. Consider a two-variable function 
$\phi(\lambda,\mu)$ possessing the property 
\be{prop} 
\phi(\lambda,\mu)=\phi(-\mu,-\lambda).
\ee
In particular one can choose an arbitrary function, depending on the
difference of the arguments. Define $f(\lambda,\mu)$ as the image of
$\phi(\lambda,\mu)$ under the integral transformation of the following
form:
\be{inttrans}
f(\lambda,\mu)=\phi(\lambda,\mu)+\stint r(\lambda,\xi)
\phi(\xi,\mu)\,d\xi.
\ee
We do not set any restrictions on the kernel 
$r(\lambda,\mu)$ and the function $\phi(\lambda,\mu)$, except the
existence of the transformation \eq{inttrans}. In other words the
integral in the r.h.s. of \eq{inttrans} must be convergent.

It is easy to see that the image of $\phi(\lambda,\mu)$ has the
property
\be{propf}
f(\lambda,\mu)-f(-\mu,-\lambda)=
\stint[r(\lambda,\nu)f(-\mu,-\nu)-r(-\mu,-\nu)f(\lambda,\nu)]\,d\nu.
\ee
In order to check the last equation one should simply substitute 
$f(\lambda,\mu)$ from \eq{inttrans} into \eq{propf} and use the
property \eq{prop}.

In fact, the Eq.\eq{propf} can be considered
as the generated relation for identities for the image of the 
transformation  \eq{inttrans}. Indeed, if the kernel $r(\lambda,\mu)$
depends on $f(\lambda,\mu)$, then the transformation \eq{inttrans} 
turns into an equation for $f$. In turn the Eq.\eq{propf} becomes an 
identity for $f$.

Consider the example related to the equation \eq{Tphase}. The formal 
solution of this equation can be given in terms of the resolvent
\be{solres}
F(\lambda,\mu)=\frac{1}{2\pi}\Phi(\lambda-\mu)
+\frac{1}{2\pi}\stint R(\lambda,\nu)\Phi(\nu-\mu)\,d\nu,
\ee
where
\be{resolv}
R(\lambda,\mu)-\frac{1}{2\pi}
\stint K(\lambda-\nu)\vartheta(\nu)R(\nu,\mu)\,d\nu
=\frac{1}{2\pi}K(\lambda-\mu)\vartheta(\mu).
\ee
The function $\Phi(\lambda-\mu)$ depends on the difference of the 
arguments and, hence, it satisfies the property \eq{prop}. Therefore
due to \eq{propf}
\be{propF}
F(\lambda,\mu)-F(-\mu,-\lambda)=
\stint[R(\lambda,\nu)F(-\mu,-\nu)-R(-\mu,-\nu)F(\lambda,\nu)]\,d\nu.
\ee
On the other hand the resolvent evidently depends on $F(\lambda,\mu)$,
since, differentiating \eq{Tphase} with respect to $\mu$, we
have
\be{Fder}
\partial_\mu F(\lambda,\mu)-\frac{1}{2\pi}
\stint K(\lambda-\nu)\vartheta(\nu)
\partial_\mu F(\nu,\mu)\,d\nu
=-\frac{1}{2\pi}K(\lambda-\mu).
\ee
Comparing with \eq{resolv}, we obtain
\be{resandF}
R(\lambda,\mu)=-\partial_\mu F(\lambda,\mu)
\vartheta(\mu).
\ee
Substituting this expression into \eq{propF}, we arrive at
\ba{premident}
&&{\dis\hspace{-2.5cm}
F(\lambda,\mu)-F(-\mu,-\lambda)=
-\stint[F(\lambda,\nu)\vartheta(-\nu)\partial_\nu F(-\mu,-\nu)}\non
&&{\dis\hspace{3.5cm}
+F(-\mu,-\nu)\vartheta(\nu)\partial_\nu F(\lambda,\nu)]\,d\nu,}
\ea
or, integrating the first term by parts
\be{ident11}
F(\lambda,\mu)-F(-\mu,-\lambda)=
\stint F(\lambda,\nu)F(-\mu,-\nu)\partial_\nu \vartheta(\nu)
\,d\nu.
\ee 
Here we have used $\vartheta(\lambda)=\vartheta(-\lambda)$.
Thus, the nonlinear identity \eq{ident1} for the scattering phase
at finite temperature is proved.

Observe that we did not use the explicit form of the kernel
$K(\lambda-\mu)$, the bare phase $\Phi(\lambda-\mu)$ and 
the Fermi weight $\vartheta(\lambda)$. Our proof was based on more 
general properties of the equation \eq{Tphase}. First, we have used 
the solvability of this equation. This provides us with the existence
of the resolvent. Second, we have used the fact, that the bare phase 
depends on the difference of the arguments. As we have seen, this 
property is not necessary and it can be replaced by more weak one:
$\Phi(\lambda,\mu)=\Phi(-\mu,-\lambda)$. Third, the most important 
property is the connection between the kernel $K(\lambda,\mu)$ and
the function $\Phi(\lambda,\mu)$: $K(\lambda,\mu)=
-\partial_\mu\Phi(\lambda,\mu)$. This allows us to express the 
resolvent of the equation \eq{Tphase} in terms of the solution of 
this equation $R(\lambda,\mu)=-\partial_\mu F(\lambda,\mu)
\vartheta(\mu)$. Finally, in order to reduce \eq{premident} to 
\eq{ident11} we have used the symmetry of the Fermi weight
$\vartheta(\lambda)=\vartheta(-\lambda)$.

The listed properties are valid for equations, describing the 
scattering phase, for a wide class of quantum integrable models
(see, for example, \cite{KBI} and references therein). These 
integral equations arise in the special limit of the Bethe equations, 
and one can say, that the nonlinear identity for the scattering
phase is the corollary of the Bethe equations.

I would like to thank S.~A.~Frolov and Yu.~M.~Zinoviev for
useful discussions. This work was supported in parts by
RFBR-96-01-166 and INTAS-93-1038.

\end{document}